\documentclass[a4paper]{article}
\usepackage[a4paper,top=3cm,bottom=2cm,left=3cm,right=3cm,marginparwidth=1.75cm]{geometry}
\usepackage{amsfonts}
\usepackage{bm}
\usepackage{extarrows}
\usepackage{amssymb}
\usepackage{color}
\usepackage[all]{xy}
\usepackage{graphicx}
\usepackage{braket}
\usepackage{amsmath}
\usepackage{appendix}
\usepackage{graphicx}
\usepackage[colorinlistoftodos]{todonotes}
\usepackage{amsthm}
\usepackage{mathrsfs}
\usepackage{amssymb}
\usepackage{accents}
\usepackage{extarrows}
\usepackage{makecell}
\usepackage{authblk}
\usepackage{url}
\usepackage{graphicx}
\usepackage{hyperref}
\usepackage{cite}
\usepackage{eufrak}
\hypersetup{colorlinks,
            citecolor=black,
            linkcolor=black,
            urlcolor=green,
            pdftex}

\def\be{\begin{equation}}
\def\ee{\end{equation}}
\def\ba{\begin{eqnarray}}
\def\ea{\end{eqnarray}}

\usepackage[english]{babel}
\usepackage[utf8x]{inputenc}
\usepackage[T1]{fontenc}

\usepackage[normalem]{ulem}

\title{The effective dynamics of weak coupling loop quantum gravity}
\author[1,2]{Gaoping Long \footnote{201731140005@mail.bnu.edu.cn}}
\author[2]{Yongge Ma \footnote{mayg@bnu.edu.cn}\thanks{corresponding author}}

\affil[1]{Department of Physics, South China University of Technology, Guangzhou 510641, China}
\affil[2]{Department of Physics, Beijing Normal University, Beijing 100875, China}

\date{}

\begin{document}

\maketitle

\begin{abstract}
By taking the limit that Newton's Gravitational constant tends to zero, the weak coupling loop quantum gravity can be formulated as a $U(1)^3$ gauge theory instead of the original $SU(2)$ gauge theory. In this paper, a parametrization of the $SU(2)$ holonomy-flux variables by the $U(1)^3$  holonomy-flux variables is introduced, and the Hamiltonian operator based on this parametrization is obtained for the weak coupling loop quantum gravity. It is shown that the effective dynamics obtained from the coherent state path integrals in $U(1)^3$ and $SU(2)$ loop quantum gravity respectively are consistent to each other in the weak coupling limit, provided that the expectation values of the Hamiltonian operators on the coherent states in these two theories coincide with their classical expressions respectively.
\end{abstract}

\section{Introduction}
Loop quantum gravity (LQG) opens a convincing approach to achieve the unification of general relativity (GR) and quantum mechanics \cite{Ashtekar2012Background,RovelliBook2,Han2005FUNDAMENTAL,thiemann2007modern,rovelli2007quantum}. The distinguished feature of LQG is its non-perturbative and background-independent construction, which predicts the discretization of spatial geometry. An interesting research topic in the field is the weak coupling limit LQG, which is given by taking the limit that the Newton's gravitational constant $\kappa$ tends to 0. This idea was firstly proposed by Smolin and further studied by Tomlin and Varadarian \cite{Smolin_1992,PhysRevD.87.044039}. The resulting weak coupling LQG is a $U(1)^3$ gauge theory instead of the original $SU(2)$ gauge theory. This $U(1)^3$ LQG theory inherits some of the core characters of the original $SU(2)$ LQG, such as the discrete spatial geometry and the polymer-like quantization scheme. It has been used as a toy model to study the faithful LQG-like representation of the constraint algebra in the weak coupling limit of Euclidean GR \cite{Lewandowski:2016lby}. The theoretical framework of the weak coupling $U(1)^3$ LQG model is also used to study the quantum field theory on curved spacetime limit of LQG \cite{Sahlmann:2002qj,Sahlmann:2002qk}.

Though the $U(1)^3$ theory has been used as a toy model of LQG, whether the predictions of the model would coincides with those of LQG is still uncertain due to the following difficulties. First, the geometric meaning of the holonomy-flux variables in the $U(1)^3$ weak coupling LQG is not clear, and hence it is hard to reveal the physical meaning of this theory. Second, the weak coupling limit of the scalar constraint in the $U(1)^3$ theory only represents certain dynamics of Euclidean gravity, which could diverge from that of the original $SU(2)$ theory of LQG, since the higher order terms of $\kappa$ has been neglected in former. Hence, the dynamics of the weak coupling limit of LQG is still an open issue. In this paper, we will concentrate on this problem and try to construct the effective dynamics from coherent state path integral of the weak coupling LQG.

To study the effective dynamics, we will first define a Hamiltonian constraint in the weak coupling $U(1)^3$ LQG. Then the effective dynamics will be constructed from the $U(1)^3$ coherent state path integral. 
More specifically, to define the Hamiltonian constraint, a parametrization of $SU(2)$ holonomy-flux variables by $U(1)^3$ holonomy-flux variables will be constructed. It will be shown that the $SU(2)$ holonomy-flux algebra can be reproduced in the $U(1)^3$ holonomy-flux phase space based on this parametrization in the weak coupling limit.  Then, the Hamiltonian constraint in the weak coupling $U(1)^3$ LQG can be given by replacing the $SU(2)$ holonomy-flux variables in the Hamiltonian constraint of the $SU(2)$ LQG with the corresponding reparametrization variables in the $U(1)^3$ holonomy-flux phase space. Also, this parametrization endows a geometric meaning to the $U(1)^3$ holonomy-flux variables. Such a parametrization is inspired by the definition of the Hamiltonian of a Fermion field in the weak coupling $U(1)^3$ LQG background, in which the $U(1)^3$ holonomies are used to construct $SU(2)$ holonomies to give the transportation of the $SU(2)$ spinors \cite{Sahlmann:2002qj}\cite{Sahlmann:2002qk}. With the Hamiltonian constraint in the weak coupling $U(1)^3$ LQG and the $U(1)^3$ complexifier coherent states, the effective dynamics will be derived following the the standard coherent state functional integral method. We will show that the equations of motion (EOMs) given by the effective dynamics of the $U(1)^3$ LQG is consistent with that of $SU(2)$ LQG in weak coupling limit up to higher order corrections of $t=\kappa\hbar/a^2$ with $a$ being a chosen unit length.

\section{Elements of LQG}
\subsection{The basic structures}
The (1+3)-dimensional Lorentzian LQG is constructed by canonically quantizing GR based on the Yang-Mills phase space with the non-vanishing Poisson bracket
 \begin{equation}
 \{A_{a}^i(x),E^{b}_j(y)\}=\kappa\beta\delta_a^b\delta^i_j\delta^{(3)}(x-y),
 \end{equation}
where the configuration and momentum are respectively the  $su(2)$-valued connection field $A_{a}^i$ and densitized triad field $E^{b}_j$ on a 3-dimensional spatial manifold $\Sigma$, $\kappa$ and $\beta$ represent the gravitational constant and Babero-Immirze parameter respectively. Here we use $i, j, k, ...$ for the internal $su(2)$ index and $a, b, c, ...$ for the spatial index. Let $q_{ab}=e_a^ie_{bi}$ be the spatial metric on $\Sigma$. The densitized triad is related to the triad $e_i^a$ by $E^{a}_{i}=\sqrt{\det(q)}e^{a}_{i}$, where $\det(q)$ denotes the determinant of $q_{ab}$. The connection can be expressed as $A_{a}^{i}=\Gamma_{a}^{i}+\beta K_{a}^{i}$, where $\Gamma_{a}^{i}$ is the Levi-Civita connection of $e_{a}^{i}$ and $K_a^i$ is related to the extrinsic curvature $K_{ab}$ by $K_a^i=K_{ab}e^b_j\delta^{ji}$. The dynamics is governed by the following Gaussian, vector and scalar constraints respectively,
\begin{equation}\label{GC}
 \mathcal{G}:=\partial_aE^{ai}+A_{aj}E^a_k\epsilon^{ijk}=0,
\end{equation}
\begin{equation}\label{VC}
 \mathcal{C}_a:=E^b_iF^i_{ab}=0,
\end{equation}
and
\begin{equation}\label{SC}
\mathcal{C}:=\frac{E^a_i E^b_j}{\det{(E)}}({\epsilon^{ij}}_kF^k_{ab}-2(1+\beta^2)K^i_{[a}K^j_{b]})=0,
\end{equation}
where $F_{ab}^i=\partial_aA_b^i-\partial_bA_a^i+\epsilon_{ijk}A_a^jA_b^k$ is the curvature of $A_a^i$.
As a totally constrained system, the physical time evolution in the Hamiltonian formulation of GR can be constructed by several deparametrization models \cite{Brown:1994py}\cite{PhysRevD.43.419}\cite{Domagala:2010bm}. In these models,  the resulting physical Hamiltonian $\mathbf{H}$ can be written as $\mathbf{H}=\int_{\Sigma}dx^3h$ with the densitized scalar field $h=h(\mathcal{C},\mathcal{C}_a)$ taking different formulations for different deparametrization models. For instance, in the Gaussian dust deparametrization model one has $h=\mathcal{C}$\cite{PhysRevD.43.419}\cite{Han_2020}.

The loop quantization of the $SU(2)$ connection formulation of GR leads to a kinematical Hilbert space $\mathcal{H}$, which can be regarded as a union of the Hilbert spaces $\mathcal{H}_\gamma=L^2((SU(2))^{|E(\gamma)|},d\mu_{\text{Haar}}^{|E(\gamma)|})$ on all possible finite graphs $\gamma$,  where $|E(\gamma)|$ denotes the number of independent edges of $\gamma$ and $d\mu_{\text{Haar}}^{|E(\gamma)|}$ denotes the product of the Haar measure on $SU(2)$. In this sense, on each given $\gamma$ there is a discrete phase space $(T^\ast SU(2))^{|E(\gamma)|}$, which is coordinatized by the basic discrete variables---holonomies and fluxes. The holonomy of $A_a^i$ along an edge $e\in\gamma$ is defined by
 \begin{equation}
h_e[A]:=\mathcal{P}\exp(\int_eA)=1+\sum_{n=1}^{\infty}\int_{0}^1dt_n\int_0^{t_n}dt_{n-1}...\int_0^{t_2} dt_1A(t_1)...A(t_n),
 \end{equation}
 where $A(t)=A_a^i(t)\dot{e}^a(t)\tau_i$, and $\tau_i=-\frac{\mathbf{i}}{2}\sigma_i$ with $\sigma_i$ being the Pauli matrices.
There are two versions for the gauge covariant flux of $E^b_j$ through the 2-face dual to edge $e\in E(\gamma)$ \cite{Thomas2001Gauge}\cite{zhang2021firstorder}. The flux in the perspective of source point of $e$ is defined by
 \begin{equation}\label{F111}
 F^i(e):=\frac{2}{\beta }\text{tr}\left(\tau^i\int_{S_e}\epsilon_{abc}h(\rho^s_e(\sigma))E^{cj}(\sigma)\tau_jh(\rho^s_e(\sigma)^{-1})\right),
 \end{equation}
 where $S_e$ is the 2-face in the dual lattice $\gamma^\ast$ of $\gamma$, $\rho^s(\sigma): [0,1]\rightarrow \Sigma$ is a path connecting the source point $s_e\in e$ to $\sigma\in S_e$ such that $\rho_e^s(\sigma): [0,\frac{1}{2}]\rightarrow e$ and $\rho_e^s(\sigma): [\frac{1}{2}, 1]\rightarrow S_e$.  Similarly, the corresponding flux in the perspective of target point of $e$ is defined by
  \begin{equation}\label{F222}
 \tilde{F}^i(e):=-\frac{2}{\beta }\text{tr}\left(\tau^i\int_{S_e}\epsilon_{abc}h(\rho^t_e(\sigma))E^{cj}(\sigma)\tau_jh(\rho^t_e(\sigma)^{-1})\right),
 \end{equation}
 where $\rho^t(\sigma): [0,1]\rightarrow \Sigma$ is a path connecting the target point $t_e\in e$ to $\sigma\in S_e$ such that $\rho_e^t(\sigma): [0,\frac{1}{2}]\rightarrow e$ and $\rho_e^t(\sigma): [\frac{1}{2}, 1]\rightarrow S_e$. It is easy to see that one has the relation
 \begin{equation}\label{F333}
 \tilde{F}^i(e)\tau_i=-h_e^{-1} {F}^i(e)\tau_ih_e.
 \end{equation}
 The non-vanishing Poisson brackets among the holonomy and fluxes read
 \begin{eqnarray}\label{hp1220}
&&\{{{h}}_e[{A}],{{F}}^i_{e'}\}= -\delta_{e,e'}{\kappa}\tau^i{{h}}_e[{A}],\quad \{{{h}}_e[{A}],\tilde{{F}}^i_{e'}\}= \delta_{e,e'}{\kappa}{{h}}_e[{A}]\tau^i,\\\nonumber
&&\{{{F}}^i_{e},{{F}}^j_{e'}\}= -\delta_{e,e'}{\kappa}{\epsilon^{ij}}_k{{F}}^k_{e'},\quad \{\tilde{{F}}^i_{e},\tilde{{F}}^j_{e'}\}= -\delta_{e,e'}{\kappa}{\epsilon^{ij}}_k\tilde{{F}}^k_{e'}.
\end{eqnarray}

The basic operators in $\mathcal{H}_\gamma$ is given by promoting the basic discrete variables as operators. The resulting holonomy and flux operators act on cylindrical functions $f_\gamma(A)=f_\gamma(h_{e_1}[A],...,h_{e_{|E(\gamma)|}}[A])$ in $\mathcal{H}_\gamma$ as
\begin{equation}
\hat{h}_e[A]f_\gamma(A)=h_e[A]f_\gamma(A),
\end{equation}
\begin{equation}
\hat{F}^i(e)f_\gamma(h_{e_1}[A],...,h_{e}[A],...,h_{e_{|E(\gamma)|}}[A]=\mathbf{i}\kappa\hbar\frac{d}{d\lambda} f_\gamma(h_{e_1}[A],...,e^{\lambda\tau^i}h_{e}[A],...,h_{e_{|E(\gamma)|}}[A],
\end{equation}
\begin{equation}
\hat{\tilde{F}}^i(e)f_\gamma(h_{e_1}[A],...,h_{e}[A],...,h_{e_{|E(\gamma)|}}[A]=-\mathbf{i}\kappa\hbar\frac{d}{d\lambda} f_\gamma(h_{e_1}[A],...,h_{e}[A]e^{\lambda\tau^i},...,h_{e_{|E(\gamma)|}}[A].
\end{equation}
Two spatial geometric operators in $H_\gamma$ are worth being mentioned here. The first one is the oriented area operator defined as $\beta \hat{F}^i(e)$ (or $\beta \hat{\tilde{F}}^i(e)$), whose module length $|\beta \hat{F}(e)|:=\sqrt{\beta^2\hat{F}^i(e)\hat{F}_i(e)}$ represents the area of the 2-face dual to $e$ and direction represents the ingoing normal direction of $S_e$ in the perspective of the source (or target) point of $e$. As a remarkable prediction of LQG, the module length and the components of the oriented area operator take respectively the following discrete spectrum \cite{Ashtekar2012Background}\cite{Han2005FUNDAMENTAL},
 \begin{equation}
 \text{Spec}(|\beta \hat{F}(e)|)=\{\beta \kappa\hbar\sqrt{j(j+1)}|j\in\frac{\mathbb{N}}{2}\},
 \end{equation}
 \begin{equation}\label{eigenp}
 \text{Spec}(\beta \hat{F}^i(e))=\{\beta \kappa\hbar m|m\in\frac{\mathbb{Z}}{2}\},\forall i=1,2,3.
 \end{equation}
 The second important spatial geometric operator is the volume operator of a compact region $R\subset \Sigma$, which is defined as
\begin{equation}\label{Vdef}
\hat{V}_R:=\sum_{v\in V(\gamma)\cap R}\hat{V}_v=\sum_{v\in V(\gamma)\cap R}\sqrt{|\hat{Q}_v|},
\end{equation}
where $V(\gamma)$ denotes the set of vertices of $\gamma$, and
\begin{equation}
\hat{Q}_v:=\frac{1}{8}(\beta )^3\sum_{\{e_I,e_J,e_K\}\subset E(\gamma)}^{e_I\cap e_J\cap e_K=v}\epsilon_{ijk}\epsilon^{IJK}\hat{F}^i(v,e_I)\hat{F}^j(v,e_J)\hat{F}^k(v,e_K),
\end{equation}
where $\epsilon^{IJK}=\text{sgn}[\det(e_I\wedge e_J\wedge e_K)]$, $\hat{F}^i(v,e)=\hat{F}^i(e)$ if $s(e)=v$ and $\hat{F}^i(v,e)=-\hat{\tilde{F}}^i(e)$ if $t(e)=v$.

The Gaussian constraint operator can be well defined in $\mathcal{H}_\gamma$ as well as in $\mathcal{H}$, which generates $SU(2)$ gauge transformations of the cylindrical functions. However, there is no operator in either $\mathcal{H}_\gamma$ or $\mathcal{H}$ corresponding to the vector constraint. To solve the diffeomorphism constraint at quantum level, one has to use the group-averaging procedure on $\mathcal{H}$ to achieve a diffeomorphism invariant Hilbert space \cite{Ashtekar2012Background}\cite{Han2005FUNDAMENTAL}. We now consider the operator in $\mathcal{H}_\gamma$ corresponding to the scalar constraint. The quantum scalar constraint is constituted by the so-called Euclidean part $\hat{\mathcal{C}}_E{[N]}$ and Lorentzian part $\hat{\mathcal{C}}_L{[N]}$ as
\begin{equation}\label{scalarcons}
\hat{\mathcal{C}}{[N]}=\hat{\mathcal{C}}_E{[N]}+(1+\beta^2)\hat{\mathcal{C}}_L{[N]},
\end{equation}
where $N$ is the smearing function. The Euclidean part is defined as
\begin{equation}
\hat{\mathcal{C}}_E{[N]}=\frac{1}{\mathbf{i}\beta \kappa\hbar}\sum_{v\in V(\gamma)}N(v)\sum^{e_I\cap e_J\cap e_K=v}_{\{e_I,e_J,e_K\}\subset E(\gamma)}\epsilon^{IJK}\text{tr}(h_{\alpha_{IJ}}h_{e_K}[\hat{V}_v,h^{-1}_{e_K}]),
\end{equation}
where $e_I,e_J, e_K$ have been re-oriented to be outgoing at $v$, $\epsilon^{IJK}=\text{sgn}[\det(e_I\wedge e_J\wedge e_K)]$, $\alpha_{IJ}$ is the minimal loop around a plaquette containing $e_I$ and $e_J$ \cite{Han_2020semiclassical,Giesel_2007}, which begins at $v$ via $e_I$ and gets back to $v$ through $e_J$. With the same notations, the Lorentzian part is given by
\begin{equation}
\hat{\mathcal{C}}_L{[N]}=\frac{-1}{2\mathbf{i}\beta^7(\kappa\hbar)^5}\sum_{v\in V(\gamma)}N(v)\sum^{e_I\cap e_J\cap e_K=v}_{\{e_I,e_J,e_K\}\subset E(\gamma)}\epsilon^{IJK}\text{tr}\left([h_{e_I},[\hat{V}_v,\hat{C}_E]]h^{-1}_{e_I} [h_{e_J},[\hat{V}_v,\hat{C}_E]]h^{-1}_{e_J}[h_{e_K},\hat{V}_v]h_{e_K}^{-1}\right).
\end{equation}

\subsection{Effective dynamics from coherent state path integral}
The dynamics of LQG can be defined by the physical Hamiltonian which is introduced by the deparametrization of GR.
 In the deparametrization models with certain dust fields, the scalar and diffeomorphism constraints are solved classically so that the theory can be described in terms of Dirac observables, since the dust reference frame provides the physical spatial coordinates and  time $\tau$. Then, the physical time evolution is generated by the physical Hamiltonian with respect to the dust field \cite{Han_2020,PhysRevD.43.419}.
 In the Gaussian dust deparametrization model \cite{PhysRevD.43.419}, the (non-graph-changing) physical Hamiltonian operator $\hat{\mathbf{H}}$ determining the quantum dynamics in $\mathcal{H}_\gamma$ can be given as
\begin{equation}\label{Hami}
\hat{\mathbf{H}}=\frac{1}{2}(\hat{\mathcal{C}}{[1]}+\hat{\mathcal{C}}{[1]}^\dagger).
\end{equation}
This operator is manifestly Hermitian and therefore admits a self-adjoint extension.
Based on this Hamiltonian operator, the effective dynamics from the coherent state path integral has been studied for a cubic graph $\gamma$ in \cite{Han_2020,Han_2020semiclassical}. We now give a brief review of this effective dynamics.

The method of coherent state path integral has been successfully applied to derive the effective dynamics in both LQG and its cosmological models \cite{Qin:2011hx,Qin:2012xh}. There are several proposals for constructing coherent states in LQG \cite{Calcinari_2020,Bianchi:2009ky,PhysRevD.104.046014,Long:2020euh,Long:2021lmd}. The most widely used one is the so-called complexifier coherent state constructed based on the heat-kernel coherent state of $SU(2)$ \cite{Thomas2001Gauge,2001Gauge,2000Gauge}. For a graph $\gamma$, the complexifier coherent state is given by
\begin{equation}
\Psi_{\gamma,{g}}^{{t}}({h})=\prod_{e\in E(\gamma)}\Psi^{t}_{g_e}(h_e)
\end{equation}
with
\begin{equation}
\Psi^{t}_{g_e}(h_e):=\sum_{j_e\in(\mathbb{Z}_+/2)\cup0}(2j_e+1)e^{-t_ej_e(j_e+1)/2}\chi_{j_e}( g_e h^{-1}_e),
\end{equation}
where ${g}=\{{g}_e\}_{e\in E(\gamma)}$, ${h}=\{{h}_e\}_{e\in E(\gamma)}$, $\chi_{j}$ is the $SU(2)$ character with spin $j$ and $t\in\mathbb{R}^+$ is a semi-classicality parameter. As a function of the holonomies $h_e=e^{\theta^i_e\tau_i}$, the coherent state is labelled by the complex coordinates $g_e\in T^\ast SU(2)\cong SL(2,\mathbb{C})$ of the discrete holonomy-flux phase space of LQG. For an edge $e$, the coordinate is the complexified
holonomy
\begin{equation}
g_e=e^{-\mathbf{i}p^i(e)\tau_i}e^{\phi^i(e)\tau_i},\quad p^i(e), \phi^i(e)\in\mathbb{R}^3
\end{equation}
where $e^{\phi^i_e\tau_i}$ parametrizes the classical holonomy variable and $p^i(e)=\frac{F^i(e)}{a^2}$ is the dimensionless flux with $a$ being a constant with the dimension of length related to the semiclassicality parameter by $t=\kappa\hbar/a^2$.   The gauge invariant coherent state is labelled by gauge equivalent class of $g(e)\sim g^{h'}(e):=h'^{-1}_{s(e)}g(e)h'_{t(e)}$ for all $e\in E(\gamma)$. The semiclassical limit is given by $t\rightarrow0$ or $\ell_P<< a$. Thanks to the overcompleteness and semiclassical properties of the coherent states, the transition amplitude between gauge invariant coherent states
can be written as the following discrete path integral formula \cite{Han_2020},
\begin{equation}
A_{g,g'}=\int dh' \langle \Psi_{\gamma,{g}}^{{t}}|\left(\exp(-\frac{\mathbf{i}}{\hbar}\Delta\tau\hat{\mathbf{H}})\right)^N |\Psi_{\gamma,{g}'^{h'}}^{{t}}\rangle=\|\Psi_{\gamma,{g}}^{{t}}\|\|
\Psi_{\gamma,{g}'}^{{t}}\|\int dh' \prod_{\imath=1}^{N+1}dg_\imath\nu[g]e^{S[g,h']/t},
\end{equation}
where the integral is taken over $N+1$ intermediate states labelled by $g_\imath\in SL(2,\mathbb{C})^{|E(\gamma)|}$ with $g_0=g'^{h'}, g_{N+2}=g$, the gauge transformation elements $h'=\{h'_v\}_{v\in V(\gamma)}\in SU(2)^{|V(\gamma)|}$ are added to ensure the $SU(2)$ gauge invariance, $\nu[g]$ is a path integral measure, $\|\Psi_{\gamma,{g}}^{{t}}\|$ is the module of the state $\Psi_{\gamma,{g}}^{{t}}(h)$, and $S[g,h']$ can be regarded as the effective action for LQG extracted from the path integral. In the continuous time limit, this action can be written as \cite{Han_2020semiclassical,zhang2021firstorder}
\begin{eqnarray}\label{action}
\nonumber\mathcal{S}[g,h']&=&\lim_{\Delta\tau=T/N\rightarrow0}S[g,h']\\\nonumber
&=&\mathbf{i}\int_{0}^Td\tau\left[\sum_{e\in E(\gamma)}X^i(\tau,e)\frac{d\phi^i(\tau,e)}{d\tau}-\frac{\kappa}{a^2}\langle \Psi_{\gamma,{g}(\tau)}^{{t}}|\hat{\mathbf{H}}| \Psi_{\gamma,{g}(\tau)}^{{t}}\rangle\right]\\
&=&\mathbf{i}\int_{0}^Td\tau\left[\sum_{e\in E(\gamma)}X^i(\tau,e)\frac{d\phi^i(\tau,e)}{d\tau}-\frac{\kappa}{a^2}\left(\mathbf{H}[\mathbf{p}(\tau), \bm{\phi}(\tau)]+\mathcal{O}(t)\right)\right],
\end{eqnarray}
where $\langle \Psi_{\gamma,{g}(\tau)}^{{t}}|\hat{\mathbf{H}}| \Psi_{\gamma,{g}(\tau)}^{{t}}\rangle=\mathbf{H}[\mathbf{p}(\tau), \bm{\phi}(\tau)]+\mathcal{O}(t)$, $\mathbf{p}=\{\mathbf{p}_e\}_{e\in\gamma}=\{p_e^i\}_{e\in\gamma}$, $\bm{\phi}=\{\bm{\phi}_e\}_{e\in\gamma}=\{\phi_e^i\}_{e\in\gamma}$ and
\begin{equation}
X^i_e=G_{ij}(\mathbf{\phi}_e)p^j_e.
\end{equation}
Here the $3\times3$ real matrix $G_{ij}(\mathbf{\phi})$ is given by
\begin{equation}
\left(\begin{array}{cccc}
    -\frac{(\phi\phi_1^2+(\phi_2^2+\phi_3^2)\sin(\phi))}{\phi^3} &    -\frac{(\phi_1\phi_2(\phi-\sin(\phi))+\phi\phi_3(\cos(\phi)-1))}{\phi^3}    & \frac{(\phi_1\phi_3(\sin(\phi)-\phi)+\phi\phi_2(\cos(\phi)-1))}{\phi^3} \\
    \frac{\phi\phi_3(\cos(\phi)-1)-\phi_1\phi_2(\phi-\sin(\phi))}{\phi^3}  &    -\frac{(\phi\phi_2^2+(\phi_1^2+\phi_3^2)\sin(\phi))}{\phi^3}   &  -\frac{(\phi_2\phi_3(\phi-\sin(\phi))+\phi\phi_1(\cos(\phi)-1))}{\phi^3}\\
     -\frac{(\phi_1\phi_3(\phi-\sin(\phi))+\phi\phi_2(\cos(\phi)-1))}{\phi^3} & \frac{(\phi_2\phi_3(\sin(\phi)-\phi)+\phi\phi_1(\cos(\phi)-1))}{\phi^3} &  -\frac{(\phi\phi_3^2+(\phi_1^2+\phi_2^2)\sin(\phi))}{\phi^3}
\end{array}\right)
\end{equation}
where $\phi_{i,e}\equiv \phi^i_e$ and $\phi=\sqrt{\phi^i_e\phi_{i,e}}$. Also, the inherent Poisson algebra of the basic variables in this effective action is
\begin{equation}
\{\phi_e^i,\phi_{e'}^j\}=\{X_e^i,X_{e'}^j\}=0,\quad  \{\phi_e^i,X_{e'}^j\}=\frac{\kappa}{a^2}\delta_{e,e'}\delta^{ij},
\end{equation}
which is equivalent to the Poisson algebra
\begin{equation}\label{poissonp}
\{\phi_e^i,\phi_{e'}^j\}=0,\quad  \{p_{e'}^i,\phi_e^j\}=\frac{\kappa}{a^2}\delta_{e,e'}U^{i}_{\  j}(\bm{\phi}),\quad \{p_e^i,p_{e'}^j\}=-\frac{\kappa}{a^2}\delta_{e,e'}{\epsilon^{ij}}_{k}p_{e}^k
\end{equation}
originated from the holonomy-flux algebra, where $U(\bm{\phi})G(\bm{\phi})^{\text{T}}=G(\bm{\phi})^{\text{T}}U(\bm{\phi})=-1_{3\times3}$ with $G(\bm{\phi})^{\text{T}}$ representing the matrix transposition of $G(\bm{\phi})$.
The variations of the action \eqref{action} with respect to $\phi^i_e$ and $X^i_e$ give the Hamiltonian equations (up to $\mathcal{O}(t)$)
\begin{equation}\label{EOMo}
\frac{d\phi^i_e}{d\tau}=\frac{\kappa}{a^2}\frac{\partial \mathbf{H}}{\partial X^i_e},\quad \frac{dX^i_e}{d\tau}=-\frac{\kappa}{a^2}\frac{\partial \mathbf{H}}{\partial \phi^i_e}.
\end{equation}
The variation of the action \eqref{action} with respect to $h'$ restricts the boundary state $\Psi_{\gamma,{g}(\tau)}^{{t}}$ by requiring that the classical discrete closure condition
 \begin{equation}
 -\sum_{e,s(e)=v}{p}^i(e)\tau_i+\sum_{e,t(e)=v}p^i(e)e^{-\phi^j_e\tau_j}\tau_ie^{\phi_e^k\tau_k}=0
 \end{equation}
 holds for $g=\{g_e\}_{e\in\gamma}$. This condition is preserved by the dynamical equations \eqref{EOMo}.

The effective EOMs \eqref{EOMo} represent the dynamics of full $SU(2)$ LQG at semiclassical level.  We can follow this approach to explore the effective dynamics of the weak coupling $U(1)^3$ LQG. To ensure that the $U(1)^3$ LQG reveals the full $SU(2)$ LQG exactly at effective level in weak coupling limit, one needs to show that the effective EOMs given by  the $U(1)^3$ LQG coincide with Eqs.\eqref{EOMo} in the weak coupling limit, by suitably relating the basic variables in the $U(1)^3$ LQG to those of $SU(2)$ LQG.
In the following two sections, we will  introduce a parametrization of the $SU(2)$ holonomy-flux variables by the $U(1)^3$  holonomy-flux variables and define a Hamiltonian operator for the weak coupling $U(1)^3$ LQG. We will show that, by identifying the geometrical meaning of the basic variables in the two versions of the parametrization, the coherent state path integrals in the $U(1)^3$ LQG and $SU(2)$ LQG can give a consistent effective dynamical description of the spacetime geometry in the weak coupling limit.

\section{The weak coupling $U(1)^3$ LQG}
\subsection{Basic structures}
The weak coupling theory of LQG is given by re-defining the connection as $\mathcal{A}_{ai}:=\kappa^{-1}A_{ai}$ and taking the limit $\kappa\rightarrow 0$, so that only the leading order terms with respect to $\kappa$ are remained in the Gaussian, vector and scalar constraints \cite{Smolin_1992,PhysRevD.87.044039,Lewandowski:2016lby}. The resulting theory is still a gauge theory with the conjugate pair $\mathcal{A}_{a}^{i}$ and $E^{b}_{j}$ satisfying
\begin{equation}
\{\mathcal{A}_{a}^{i}(x),E^{b}_{j}(y)\}=\beta\delta_a^b\delta^i_j\delta^{(3)}(x-y).
\end{equation}
The Gaussian constraint reduces to $\underline{\mathcal{G}}^i:=\partial_aE^{ai}$, which generates the Abilean $U(1)^3$ transformations. The reduced vector and scalar constraints are given by \cite{PhysRevD.87.044039,Lewandowski:2016lby}
\begin{equation}
 \underline{\mathcal{C}}_a:=\kappa E^b_i\underline{F}^i_{ab}
\end{equation}
and
\begin{equation}\label{Uscalar}
\underline{\mathcal{C}}:=\kappa\frac{E^a_i E^b_j}{\det{(E)}}{\epsilon^{ij}}_k\underline{F}^k_{ab}
\end{equation}
respectively, where $\underline{F}_{ab}^i=\partial_a\mathcal{A}_b^i-\partial_b\mathcal{A}_a^i$ is the curvature of $\mathcal{A}_a^i$. Here we note that the scalar constraint only contains the Euclidean part as the treatments in \cite{PhysRevD.87.044039,Lewandowski:2016lby}.

  The kinematic Hilbert space $\mathcal{K}$ of the weak coupling theory follows from the representation of the holonomy-flux algebra as in the standard LQG. Now, the holonomy is defined with an oriented curve $e\in\Sigma$ as
\begin{equation}
\underline{h}_{e}^i[\mathcal{A}]\equiv e^{\mathbf{i}\kappa\int_{e}\mathcal{A}_a^idx^a}.
\end{equation}
One way to identify a basis of the kinematic Hilbert space is to define the so-called charged holonomy $\underline{h}_{e,\vec{q}}[\mathcal{A}]$ with a triple of integer charges $\{q^i\}\equiv\vec{q}$ as
\begin{equation}
\underline{h}_{e,\vec{q}}[\mathcal{A}]\equiv e^{\mathbf{i}\kappa q_i\int_{e}\mathcal{A}_a^idx^a}.
\end{equation}
Given a closed, oriented graph $\gamma$ consisting of a set of edges $\{e_I\}$ meeting only at their end points, called the vertices, one may assign $\{\vec{q}_I\}$ to the edge $e_I\in\gamma$ and thereby
define the graph holonomy $\underline{h}_{\gamma,\{\vec{q}_I\}}$ as
\begin{equation}\label{grho}
\underline{h}_{\gamma,\{\vec{q}_I\}}[\mathcal{A}]\equiv \prod_{I}\underline{h}_{e_I,\vec{q}_I}[\mathcal{A}].
\end{equation}
Note that, as in $SU(2)$ LQG, the kinematical Hilbert space $\mathcal{K}$ can be regarded as a union of the graph-dependent Hilbert  spaces $\mathcal{K}_{\gamma}\equiv L^2\left((U(1)^{3})^{|E(\gamma)|}, d\mu_{\text{Haar}}^{|E(\gamma)|}\right)$ on all possible graphs $\gamma$ with each $U(1)^3$ associated to an edge being thought as its holonomies. Here
$L^2\left((U(1)^{3})^{|E(\gamma)|}\right)$ is the space of square-integrable functions on $(U(1)^{3})^{|E(\gamma)|}$, and $d\mu_{\text{Haar}}^{|E(\gamma)|}$ denotes the product of the Haar measure on $U(1)^3$.
A graph holonomy \eqref{grho} is local $U(1)^3$ invariant and thus a solution to the Gaussian constraint, if and only if the full set of edges $\{e_{I_v}\}$ sharing any vertex $v\in\gamma$ always satisfy the charge neutrality
\begin{equation}\label{closure}
\sum_{I_v}\text{sgn}_{I_v}q^i_{I_v}=0
\end{equation}
for all $i$, where $\text{sgn}_{i_v}$ is a positive or negative sign if the edge $e_{I_v}$ is out-going or in-going for $v$. We now define a locally $U(1)^3$ invariant charge network state, denoted as $c\equiv c(\gamma,\{\vec{q}_I\})$, to be a kinematic quantum state with a wave functional $h_c$
given by its associated graph holonomy satisfying \eqref{closure}. The $U(1)^3$ invariant kinematic Hilbert space $\mathcal{K}_{\text{inv}}\equiv\text{Span}\{|c\rangle\}$ is spanned by the basis of all the distinct charge network states and equipped with the inner product
\begin{equation}\label{inner}
\langle c| c' \rangle=\delta_{c,c'}.
\end{equation}
 Note that the labeling $(\gamma,\{\vec{q}_I\})$ to the charge network states is not unique, since one can always artificially change $\gamma$ into $\gamma'$ by adding trivial vertices and edges. To avoid this redundancy we will always label a charge network state by the corresponding oriented graph with the minimal number of edges.
The $U(1)^3$ invariant flux variables for $\underline{E}^{ai}$ is defined over an oriented 2-surface. In the case that the 2-surface $S_e$ is dual to an edge $e$ of $\gamma$, the flux is given by
\begin{equation}
\underline{F}^i(e)\equiv\frac{1}{\beta } \int_{S_e}\epsilon_{abc}{\underline{E}}^{ai}d\sigma^{b}\wedge d\sigma^{c}.
\end{equation}
The holonomy-flux Poisson bracket reads
\begin{equation}\label{originalU(1)3}
\{\underline{h}_{\gamma,\{\vec{q}_I\}},\underline{F}^i(e)\}=\sum_{e'\in \gamma(S_e)}\frac{\mathbf{i}\kappa }{2}\epsilon(e',S_e)q^i_{e'}h_{\gamma,\{\vec{q}_I\}},
\end{equation}
where $\epsilon(e',S_e)$ is the sign of the relative orientation between the given $e'$ and $S$ if they are dual to each other, and is zero otherwise, $\gamma(S_e)$ has been adapted to $S_e$ by adding pseudo vertices such that they only intersect at the vertices of the former.
In the Hilbert space $\mathcal{K}_{\gamma}$, a holonomy operator acts as a multiplicative operator.
 A flux operator then acts as a differential operator such that
\begin{equation}
\underline{\hat{F}}^i(e)\cdot\underline{h}_{\gamma,\{\vec{q}_I\}}[\mathcal{A}]=\sum_{e'\in \gamma(S_e)}\hbar\frac{\kappa}{2}\epsilon(e',S_e)q^i_{e'}\underline{h}_{\gamma,\{\vec{q}_I\}}[\mathcal{A}].
\end{equation}

The Hilbert space $\mathcal{K}$ of this $U(1)^3$ theory also has a coherent state basis. For the given graph $\gamma$, the heat kernel coherent states in this theory are given by
 \begin{equation}
\underline{\Psi}^{{t}}_{\gamma,{\underline{g}}}({\underline{h}})=\prod_{e\in\gamma} \underline{\Psi}^{t}_{\underline{g}(e)}(\underline{h}(e))
\end{equation}
where ${\underline{h}}:=\{\underline{h}(e)|e\in\gamma\}$, and ${\underline{g}}:=\{\underline{g}(e)|e\in\gamma\}$ coordinatizes the holonomy-flux phase space $(T^\ast U(1)^3)^{|E(\gamma)|}$, and $\underline{\Psi}^{t}_{\underline{g}(e)}(\underline{h}(e))$ denotes the heat kernel coherent states for $U(1)^3$ defined by
\begin{equation}
\underline{\Psi}^{t}_{\underline{g}(e)}(\underline{h}(e)):=\prod_{i\in\{1,2,3\}} \sum_{n_i=-\infty}^{\infty}e^{-\frac{t}{2}n_i^2}e^{\mathbf{i}n_i(\underline{\phi}_i(e)-\underline{\theta}_i(e) )}e^{-n_i \underline{X}_i(e)}
\end{equation}
such that $\underline{h}(e)=e^{\mathbf{i}\sum_{i}\underline{\theta}_{i}(e)}$ and $\underline{g}(e)=e^{\mathbf{i}\sum_{i}(\underline{\phi}_i(e)+\mathbf{i}\underline{X}_i(e) )}$ with $\underline{X}_i(e)
:=\frac{\underline{F}_i(e) }{a^2}$ being the dimensionless flux in the $U(1)^3$ theory.
\subsection{The issue of geometric interpretation}
The weak coupling $U(1)^3$ LQG theory captures the core characters of the full $SU(2)$ LQG with the polymer quantization scheme. The oriented area operator in the weak coupling $U(1)^3$ LQG can be defined by the flux operators similarly to that in full $SU(2)$ LQG.  Then it is easy to see that this area operator $|\beta \hat{\underline{F}}(e)|$ takes the discrete eigenvalues as
\begin{equation}
 \text{Spec}(|\beta \hat{\underline{F}}(e)|):=\text{Spec}(\sqrt{\beta^2 \hat{\underline{F}}^i(e)\hat{\underline{F}}_i(e)})=\{\beta \kappa\hbar\sqrt{\sum_{i\in\{1,2,3\}}n_i^2}|n_i\in{\mathbb{N}}\},
 \end{equation}
 due to
 \begin{equation}\label{eigenp2}
 \text{Spec}(\beta \hat{\underline{F}}^i(e))=\{\beta \kappa\hbar m|m\in\mathbb{Z}\},\forall i=1,2,3.
 \end{equation}
Note that the fluxes $F^i(e)$ and $\tilde{F}^i(e)$ represent the oriented areas of the 2-faces
dual to $e$ in the perspective of the source or target point of $e$ respectively. Recall that, in full $SU(2)$ LQG, the holonomy along an edge $e$ parallelly transports the flux from the source point to the target point of $e$ as $F_e=-h_e\tilde{F}_eh_e^{-1}$. Following the twisted geometric explanation, the Levi-Civita connection $\Gamma_e$ in the expression of the $SU(2)$ connection contributes two degrees of freedom to $h_e$ which transform $\tilde{F}_e$, and the extrinsic curvature one-form $K_e$ contributes one degree of freedom to $h_e$ which keeps $\tilde{F}_e$ invariant \cite{PhysRevD.82.084040}. However, in the weak coupling $U(1)^3$ LQG, the $U(1)^3$ holonomy along an edge $e$, which still contains 3 degrees of freedom, does not generate any transportation of the flux along $e$.
Thus, the three degrees of freedom in $U(1)^3$ holonomy can not be interpreted as a combination of the intrinsic and extrinsic curvature. In fact, the transportation of the flux along an edge in full $SU(2)$ LQG is related to the next leading order term with respect to $\kappa$ in the original Gaussian constraint $\mathcal{G}^i=\partial_aE^{ai}+\kappa\epsilon_{ijk}\mathcal{A}_a^jE^{ak}$, which is neglected in the weak coupling $U(1)^3$ theory by taking the limit $\kappa\rightarrow 0$. This limit indicates that the original $su(2)$-valued connection $A_{ai}=\kappa\mathcal{A}_{ai}$ is small so that the corresponding $SU(2)$ holonomy is almost identity and which leads to $F_e=-\tilde{F}_e$. This indicates that the weak coupling $U(1)^3$ LQG would correspond to the almost vanishing spatial curvature case of full $SU(2)$  LQG.

Whether there are higher order terms with respect to $\kappa$ may lead to difference in the dynamics. In the weak coupling $U(1)^3$ theory, one proposal to construct the scalar constraint operator is to regularize and quantize the scalar constraint \eqref{Uscalar}, so that it adds some non-degenerate vertices to the charge network state \cite{PhysRevD.87.044039}, rather than attach small loops based at the original vertices as in the usual construction of full $SU(2)$ LQG \cite{Ashtekar2012Background,Han2005FUNDAMENTAL}. However, there is no guarantee that such dynamical construction of the weak coupling $U(1)^3$ LQG can be generalized to that of full $SU(2)$ LQG.
  To employ the weak coupling $U(1)^3$ theory as a toy model to study the dynamical construction of the full $SU(2)$ LQG, one treatment is to replace the $SU(2)$ holonomy-flux operators in the scalar constraint operator of $SU(2)$ LQG by the corresponding $U(1)^3$ holonomy-flux operators \cite{Giesel_2007,Giesel_20072}. However, such scheme is only valid for the Euclidean part of the constraint but not for the Lorentzian part.
Actually, to study the weak coupling limit of the dynamics of full $SU(2)$ theory, one should not use the weak coupling limit of the constraints, since the higher-order terms with respect to $\kappa$ in those constraints may become lower-order after taking the Poisson brackets with the basic variables in the $U(1)^3$ theory. Rather, the weak coupling limit theory at the dynamical level should be given by taking the weak coupling limit of the Poisson brackets of the constraints and basic variables in the original $SU(2)$ theory.

 To deal with the above issues in the weak coupling $U(1)^3$ LQG, we are going to relate the full $SU(2)$ LQG theory and  the weak coupling $U(1)^3$ theory through reparameterizing the $SU(2)$ holonomy-flux variables by the $U(1)^3$ holonomy-flux variables. By such a parametrization, the $U(1)^3$ holonomy-flux variables can be endowed with certain geometric meanings. Also, the Gaussian constraint, vector constraint and scalar constraint in the weak coupling $U(1)^3$ LQG can be obtained by replacing the corresponding variables in the corresponding constraints of the $SU(2)$ theory.
\subsection{Re-parametrization}
 We will show in this subsection that a parametrization of the $SU(2)$ holonomy-flux variables by the $U(1)^3$ holonomy-flux variables can be realized by defining some new variables in the $U(1)^3$ holonomy-flux phase space. By this parametrization the Poisson structure of the $SU(2)$ holonomy-flux variables can be faithfully inherited in the weak coupling limit, which is consistent with the original setting of the $U(1)^3$ LQG.

For a given graph $\gamma$, the discrete phase space $T^\ast SU(2)$ and $T^\ast U(1)^3$ of the $SU(2)$ theory and $U(1)^3$ theory have the same dimensionality. Hence it is reasonable to construct a re-parametrization of the $SU(2)$ holonomy-flux by the $U(1)^3$ holonomy-flux variables.
Taking account of the expressions \eqref{F111} and \eqref{F222} of the covariant fluxes, the re-parametrization can be given by
\begin{eqnarray}\label{repa}
\underline{h}_{e}^i[\mathcal{A}]&\mapsto& h_{e}[A]:\quad \quad \quad h_{e}[A]\equiv\underline{\tilde{h}}_{e}[\mathcal{A}]:=\exp\left(\frac{\sum_{i}\left(\underline{h}_{e}^i[\mathcal{A}] -(\underline{h}_{e}^i[\mathcal{A}])^{-1}\right)\tau_i}{2\mathbf{i}}\right),\\\nonumber
\underline{X}^i_e&\mapsto&{p}^i_e:\quad \quad \quad   \quad \quad   \quad {p}^i_e\tau_i\equiv\underline{p}^i_e\tau_i:=-\underline{\tilde{h}}_{e}[\mathcal{A}/2]\tau_i \underline{\tilde{h}}^{-1}_{e}[\mathcal{A}/2]{\underline{X}}^i_e
\end{eqnarray}
with $\underline{\tilde{h}}_{e}[\mathcal{A}/2]:=\exp\left(\frac{\sum_{i}\left(\underline{h}_{e}^i[\mathcal{A}] -(\underline{h}_{e}^i[\mathcal{A}])^{-1}\right)\tau_i}{4\mathbf{i}}\right)$, where $\exp$ denotes the exponential map of $su(2)$.
Then, by defining $\tilde{\underline{p}}^i_e\tau_i:=\underline{\tilde{h}}^{-1}_{e}[\mathcal{A}/2]\tau_i \underline{\tilde{h}}_{e}[\mathcal{A}/2]{\underline{X}}^i_e$ in the weak coupling $U(1)^3$ theory, we have the relation similar to \eqref{F333} for the two fluxes of different perspectives as
\begin{equation}
\tilde{\underline{p}}^i_e\tau_i=-\underline{\tilde{h}}^{-1}_{e}[\mathcal{A}]{\underline{p}}^i_e\tau_i\underline{\tilde{h}}_{e}[\mathcal{A}].
\end{equation}
Now let us check whether the Poisson algebra of the $SU(2)$ holonomy and fluxes defined by \eqref{repa} in the $U(1)^3$ phase space coincides with that in the $SU(2)$ phase space in certain limit. With the Poisson bracket in the $U(1)^3$ theory, we obtain
\begin{equation}\label{hp0}
\{\underline{\tilde{h}}_{e}[\mathcal{A}],\underline{\tilde{h}}_{e'}[\mathcal{A}]\}=0,
\end{equation}
\begin{equation}\label{hp10}
\{\underline{\tilde{h}}_e[\mathcal{A}],{\underline{p}}^i_{e'}\}= \delta_{e,e'}\frac{2\mathbf{i}\kappa}{a^2}\sum_{j}\text{tr}(\tau^i\underline{\tilde{h}}_{e}[\mathcal{A}/2] \tau_j\underline{\tilde{h}}^{-1}_e[\mathcal{A}/2]) \underline{h}_e^j[\mathcal{A}]\frac{\delta \underline{\tilde{h}}_{e}[\mathcal{A}]}{\delta \underline{h}_e^j[\mathcal{A}]},
\end{equation}
\begin{equation}\label{hp2}
\{\underline{\tilde{h}}_e[\mathcal{A}],\tilde{\underline{p}}^i_{e'}\}= - \delta_{e,e'}\frac{2\mathbf{i}\kappa}{a^2}\sum_{j}\text{tr}(\tau^i\underline{\tilde{h}}^{-1}_{e}[\mathcal{A}/2] \tau_j\underline{\tilde{h}}_e[\mathcal{A}/2]) \underline{h}_e^j[\mathcal{A}]\frac{\delta \underline{\tilde{h}}_{e}[\mathcal{A}]}{\delta \underline{h}_e^j[\mathcal{A}]},
\end{equation}
and
\begin{eqnarray}\label{hp3}
&&\{\tilde{\underline{p}}^i_{e},\tilde{\underline{p}}^j_{e'}\}\\\nonumber
&=& -2\delta_{e,e'}\text{tr}\left(\tau^i\{\underline{\tilde{h}}^{-1}_e[\mathcal{A}/2], \tilde{\underline{p}}^j_{e}\}\tau_k\underline{\tilde{h}}_e[\mathcal{A}/2] \right)\underline{X}^k_{e}-2\delta_{e,e'}\text{tr}\left(\tau^i\underline{\tilde{h}}^{-1}_e[\mathcal{A}/2]\tau_k \{\underline{\tilde{h}}_e[\mathcal{A}/2] ,\tilde{\underline{p}}^j_{e}\}\right)\underline{X}^k_{e}\\\nonumber
&&+ 2\delta_{e,e'}\text{tr}\left(\tau^i\underline{\tilde{h}}^{-1}_e[\mathcal{A}/2]\tau_k\underline{\tilde{h}}_e[\mathcal{A}/2] \right)\left(2\text{tr}\left(\tau^j\{\underline{\tilde{h}}^{-1}_e[\mathcal{A}/2], {\underline{X}}^k_{e}\}\tau_l\underline{\tilde{h}}_e[\mathcal{A}/2] \right)\underline{X}^l_{e}\right)\\\nonumber
&&+ 2\delta_{e,e'}\text{tr}\left(\tau^i\underline{\tilde{h}}^{-1}_e[\mathcal{A}/2]\tau_k\underline{\tilde{h}}_e[\mathcal{A}/2] \right)\left(2\text{tr}\left(\tau^j\underline{\tilde{h}}^{-1}_e[\mathcal{A}/2]\tau_l \{\underline{\tilde{h}}_e[\mathcal{A}/2] ,{\underline{X}}^k_{e}\}\right)\underline{X}^l_{e}\right),
\end{eqnarray}
wherein
\begin{equation}\label{hp1}
\{\underline{\tilde{h}}_e[\mathcal{A}],{\underline{X}}^i_{e'}\}=\delta_{e,e'}\frac{\mathbf{i}\kappa}{a^2} \underline{h}_e^i[\mathcal{A}]\frac{\delta \underline{\tilde{h}}_{e}[\mathcal{A}]}{\delta \underline{h}_e^i[\mathcal{A}]},\quad (\text{No summation over}\ i).
\end{equation}
Thus, the Poisson algebra of the $SU(2)$ holonomy and fluxes defined by \eqref{repa} in the $U(1)^3$ phase space does not coincide with that in the $SU(2)$ phase space in general. Consider the $U(1)^3$ holonomy $\underline{h}^i(e)=e^{\mathbf{i}\underline{\phi}^i_e}$ and the $SU(2)$ holonomy $h(e)=e^{\phi_e^i\tau_i}$. Then it is easy to check that in the weak coupling limit given by small $\underline{\phi}^i_e=\phi_e^i$, one has $h_{e}[A]=\underline{\tilde{h}}_{e}[\mathcal{A}]$ at the leading order of $\underline{\phi}^i_e=\phi_e^i$. Further, we have 
 the Poisson algebras
\begin{eqnarray}\label{hp220}
&&\{{{h}}_e[{A}],{{p}}^i_{e'}\}= -\delta_{e,e'}\frac{\kappa}{a^2}\tau^i,\quad \{{{h}}_e[{A}],\tilde{{p}}^i_{e'}\}= \delta_{e,e'}\frac{\kappa}{a^2}\tau^i,\\\nonumber
&&\{{{p}}^i_{e},{{p}}^j_{e'}\}= -\delta_{e,e'}\frac{\kappa}{a^2}{\epsilon^{ij}}_k{{p}}^k_{e'},\quad \{\tilde{{p}}^i_{e},\tilde{{p}}^j_{e'}\}= -\delta_{e,e'}\frac{\kappa}{a^2}{\epsilon^{ij}}_k\tilde{{p}}^k_{e'}
\end{eqnarray}
in $SU(2)$ phase space and
\begin{eqnarray}\label{hp33}
&&\{\underline{\tilde{h}}_e[\mathcal{A}],{\underline{p}}^i_{e'}\}= -\delta_{e,e'}\frac{\kappa}{a^2}\tau^i,\quad \{\underline{\tilde{h}}_e[\mathcal{A}],\tilde{\underline{p}}^i_{e'}\}= \delta_{e,e'}\frac{\kappa}{a^2}\tau^i,\\\nonumber
&&\{{\underline{p}}^i_{e},{\underline{p}}^j_{e'}\}= -\delta_{e,e'}\frac{\kappa}{a^2}{\epsilon^{ij}}_k{\underline{p}}^k_{e'},\quad\{\tilde{\underline{p}}^i_{e},\tilde{\underline{p}}^j_{e'}\}= -\delta_{e,e'}\frac{\kappa}{a^2}{\epsilon^{ij}}_k\tilde{\underline{p}}^k_{e'}
\end{eqnarray}
in $U(1)^3$ phase space at the leading order of $\underline{\phi}^i_e=\phi_e^i$. Therefore, the Poisson algebra of the $SU(2)$ holonomy and fluxes defined by \eqref{repa} in the $U(1)^3$ phase space does coincide with that in the $SU(2)$ phase space in the weak coupling limit.
Moreover, the parametrization \eqref{repa} is commutative with the re-orientation of the edges as
\begin{equation}
\underline{\tilde{h}}^{-1}_{e}[\mathcal{A}]=\underline{\tilde{h}}_{e^{-1}}[\mathcal{A}], \quad\tilde{\underline{p}}^i_{e^{-1}}={\underline{p}}^i_{e},\quad {\underline{p}}^i_{e^{-1}}=\tilde{\underline{p}}^i_{e}.
\end{equation}
Thus, the variables $\underline{\tilde{h}}_{e}[\mathcal{A}]$, ${\underline{p}}^i_{e}$ and $\tilde{\underline{p}}^i_e$ in $U(1)^3$ theory inherit the explicit structure of the corresponding variables of $SU(2)$ LQG.

By construction, the variables $\underline{\tilde{h}}_{e}[\mathcal{A}]$, ${\underline{p}}^i_{e}$ and $\tilde{\underline{p}}^i_e$ in $U(1)^3$ theory can be directly quantized as
\begin{eqnarray}\label{reop}
\hat{\underline{\tilde{h}}}_{e}[\mathcal{A}]&:=& \exp\left(\frac{1}{2\mathbf{i}}\sum_{i}\left(\underline{\hat{h}}_{e}^i[\mathcal{A}]-(\underline{\hat{h}}_{e}^i[\mathcal{A}])^{-1}\right)\tau_i\right),\\\nonumber
\hat{{\underline{p}}}^j_e&:=& \text{tr}(\tau^j \hat{\underline{\tilde{h}}}_{e}[\mathcal{A}/2]\tau_i\hat{\underline{\tilde{h}}}^{-1}_{e}[\mathcal{A}/2]) \underline{\hat{X}}^i_e+\underline{\hat{X}}^i_e\text{tr}(\tau^j \hat{\underline{\tilde{h}}}_{e}[\mathcal{A}/2]\tau_i\hat{\underline{\tilde{h}}}^{-1}_{e}[\mathcal{A}/2])\\\nonumber
\hat{\tilde{\underline{p}}}^j_e&:=& -\text{tr}(\tau^j \hat{\underline{\tilde{h}}}_{e}^{-1}[\mathcal{A}/2]\tau_i\hat{\underline{\tilde{h}}}_{e}[\mathcal{A}/2]) \underline{\hat{X}}^i_e-\underline{\hat{X}}^i_e\text{tr}(\tau^j \hat{\underline{\tilde{h}}}_{e}^{-1}[\mathcal{A}/2]\tau_i\hat{\underline{\tilde{h}}}_{e}[\mathcal{A}/2]).
\end{eqnarray}
The operators $\hat{{\underline{p}}}^j_e$ and $\hat{\tilde{\underline{p}}}^j_e$ are symmetric and hence admit self-adjoint extensions.
Based on the parametrization \eqref{repa}, we can replace the basic operators in the $SU(2)$ LQG by those of the $U(1)^3$ theory in the weak coupling limit as
\begin{equation}
\hat{h}_{e}[A]\leftrightarrow\hat{\underline{\tilde{h}}}_{e}[\mathcal{A}],\quad
{\hat{p}}^i_e\leftrightarrow\hat{\underline{p}}^i_e\quad\hat{\tilde{p}}^i_e\leftrightarrow\hat{\tilde{\underline{p}}}^i_e.
\end{equation}
For instance, the corresponding volume operator $\hat{\underline{V}}_R$ in the $U(1)^3$ theory can be easily constructed by replacing ${\hat{p}}^i_e$ and ${\hat{\tilde{p}}}^i_e$ by ${\hat{\underline{p}}}^i_e$ and ${\hat{\tilde{\underline{p}}}}^i_e$ respectively in the definition \eqref{Vdef} of $\hat{V}_R$ in LQG.

Recall that the discrete version of the Gaussian constraint in $SU(2)$ LQG reads
\begin{equation}
\sum_{e,s(e)=v}\hat{p}^i_e+\sum_{e,t(e)=v}\hat{\tilde{p}}^i_e=0.
\end{equation}
Then, the corresponding discrete ``Gaussian constraint'' in the weak coupling $U(1)^3$ LQG can be given directly as
\begin{equation}
\sum_{e,s(e)=v}\hat{\underline{p}}^i_e+\sum_{e,t(e)=v}\hat{\tilde{\underline{p}}}^i_e=0,
\end{equation}
though this ``Gaussian constraint'' does not generate the $U(1)^3$ gauge transformations. In fact, it is just the closure condition for the 3-polyhedra described by its oriented areas \cite{PhysRevD.83.044035,PhysRevD.82.084040,Long:2020agv,PhysRevD.103.086016}.
Similarly, the quantum scalar constraint $\hat{\underline{\mathcal{C}}}{[N]}$ in the weak coupling theory corresponding to \eqref{scalarcons} is also constituted by the Euclidean part $\hat{\underline{\mathcal{C}}}_E{[N]}$ and Lorentzian part $\hat{\underline{\mathcal{C}}}_L{[N]}$ as
\begin{equation}
\hat{\underline{\mathcal{C}}}{[N]}=\hat{\underline{\mathcal{C}}}_E{[N]}+(1+\beta^2)\hat{\underline{\mathcal{C}}}_L{[N]}.
\end{equation}
 By acting on a cylindrical function over $\gamma$, one version of the Euclidean scalar constraint can be written as
\begin{equation}
\hat{\underline{\mathcal{C}}}_E{[N]}=\frac{1}{\mathbf{i}\beta \kappa\hbar}\sum_{v\in V(\gamma)}N(v)\sum_{e_I,e_J,e_K\ \text{at}\ v}\epsilon^{IJK}\text{tr}(\underline{h}_{\alpha_{IJ}}\underline{h}_{e_K}[\hat{\underline{V}}_v,\underline{h}^{-1}_{e_K}]),
\end{equation}
where $e_I,e_J, e_K$ are re-oriented to be outgoing at $v$, $\epsilon^{IJK}=\text{sgn}[\det(e_I\wedge e_J\wedge e_K)]$, $\alpha_{IJ}$ is the minimal loop around a plaquette containing $e_I$ and $e_J$, which begins at $v$ via $e_I$ and gets back to $v$ through $e_J$. With the same notations, the Lorentzian part $\hat{\underline{\mathcal{C}}}_L{[N]}$ is given by
\begin{equation}
\hat{\underline{\mathcal{C}}}_L{[N]}=\frac{-1}{2\mathbf{i}\beta^7(\kappa\hbar)^5}\sum_{v}N(v)\sum_{e_I,e_J,e_K\ \text{at}\ v}\epsilon^{IJK}\text{tr}\left([\underline{h}_{e_I},[\hat{\underline{V}}_v,\hat{\underline{C}}_E]]\underline{h}^{-1}_{e_I} [\underline{h}_{e_J},[\hat{\underline{V}}_v,\hat{\underline{C}}_E]]\underline{h}^{-1}_{e_J} [\underline{h}_{e_K},\hat{\underline{V}}_v]\underline{h}_{e_K}^{-1}\right).
\end{equation}
In the deparametrization formalism, the physical Hamiltonian corresponding to \eqref{Hami} reads
$\hat{\underline{\mathbf{H}}}=\frac{1}{2}(\hat{\underline{\mathcal{C}}}{[1]}+\hat{\underline{\mathcal{C}}}{[1]}^\dagger)$
  in the weak coupling theory. Thus, it
 is manifestly Hermitian and therefore admits a self-adjoint extension.
Such a physical Hamiltonian operator in the weak coupling $U(1)^3$ LQG keeps the full expression of original physical Hamiltonian in full $SU(2)$ LQG.
It is reasonable to expect that $\hat{\underline{\mathbf{H}}}$ determine the evolution which represents that of the full $SU(2)$ LQG in the weak coupling limit.
\section{Coherent state path integral of $U(1)^3$ LQG}
\subsection{Effective action and equations of motion}
With the physical Hamiltonian operator $\hat{\underline{\mathbf{H}}}$ in weak coupling $U(1)^3$ LQG, we may derive its effective dynamics based on the coherent state path integral.
The heat kernel coherent state for $U(1)^3$ gauge theory can be written as \cite{Thomas2001Gauge,2001Gauge}
\begin{equation}
\underline{\Psi}^{t}_{\underline{g}(e)}(\underline{h}(e)):=\prod_{i\in\{1,2,3\}} \sum_{n_i=-\infty}^{\infty}e^{-\frac{t}{2}n_i^2}e^{\mathbf{i}n_i(\underline{\phi}_i(e) -\underline{\theta}_i(e))}e^{-n_i \underline{X}_i(e)}
\end{equation}
at every edge $e$.
Its normalized version reads
\begin{equation}
\tilde{\underline{\Psi}}^{t}_{\underline{g}(e)}(\underline{h}(e))=\frac{\underline{\Psi}^{t}_{\underline{g}(e)}(\underline{h}(e))} {\|{\underline{\Psi}}^{t}_{\underline{g}(e)}\|}.
\end{equation}
It is important that the normalized coherent states form an over-complete basis in $\underline{\mathcal{H}}(e)=L^2(U(1)^3)$ as
\begin{equation}
\int_{G^{\mathbb{C}}}d\underline{g}(e)|\tilde{\underline{\Psi}}^{t}_{\underline{g}(e)}\rangle\langle \tilde{\underline{\Psi}}^{t}_{\underline{g}(e)}| =1_{\underline{\mathcal{H}}(e)},
\end{equation}
where
\begin{equation}
 d\underline{g}(e)=\frac{c}{t^3}\prod_{i}d\underline{\theta}_i(e)d\underline{X}_i(e),\quad\text{with}\  c=1+\mathcal{O}(t^\infty).
\end{equation}
The overlap amplitude between two coherent states reads
\begin{eqnarray}\label{overlap}
&&\langle\tilde{\underline{\Psi}}^t_{\underline{g}_2(e)},\tilde{\underline{\Psi}}^t_{\underline{g}_1(e)}\rangle \\\nonumber &=&\prod_{i\in\{1,2,3\}}\frac{e^{-\frac{\frac{1}{2}(\underline{\phi}_2^i(e)-\underline{\phi}_1^i(e))^2}{2t}} e^{-\frac{\frac{1}{2}(\underline{X}_2^i(e)-\underline{X}_1^i(e))^2}{2t}} e^{\frac{\mathbf{i}(\underline{\phi}_2^i(e)-\underline{\phi}_1^i(e))(\underline{X}_2^i(e)+\underline{X}_1^i(e))}{2t}} \sum_nf_n(\underline{\phi}_2^i(e), \underline{\phi}_1^i(e),\underline{X}_2^i(e),\underline{X}_1^i(e))}{\sqrt{D^t_{\underline{X}_2^i(e)}D^t_{\underline{X}_1^i(e)}}}
\end{eqnarray}
where $D^t_{\underline{X}}=\sum_ne^{-\frac{\pi n^2-2\mathbf{i}\pi n \underline{X}}{t}}$ and $f_n(\underline{\phi}_2^i, \underline{\phi}_1^i,\underline{X}_2^i,\underline{X}_1^i)=e^{-\frac{2\pi^2n^2-2\mathbf{i} \pi n(\underline{X}_2^i+\underline{X}_1^i) -2\pi n(\underline{\phi}_2^i-\underline{\phi}_1^i)}{2t}}$. Note that there exist constants $K_t$ and $\tilde{K}_t$ and $\tilde{K}'_t$ (independent of $\underline{g}_1(e)$ and $\underline{g}_2(e)$), decaying exponentially fast to $0$ as $t\rightarrow 0$, such that $1+K_t\geq|D_{\underline{X}(e)}^t|\geq1-K_t$ and
\begin{equation}
{(1+\tilde{K}_t)}\leq|\sum_{n}f_n(\underline{\phi}_2^i(e), \underline{\phi}_1^i(e),\underline{X}_2^i(e),\underline{X}_1^i(e))|\leq{(1+\tilde{K}'_t)}, \quad \text{for} \ |\underline{\phi}_2^i(e)-\underline{\phi}_1^i(e)|<<1.
\end{equation}
Also, the factor $e^{-\frac{\frac{1}{2}(\underline{\phi}_2^i(e)-\underline{\phi}_1^i(e))^2}{2t}}$ in \eqref{overlap} indicates that  this overlap amplitude is only non-vanishing for $|\underline{\phi}_2^i(e)-\underline{\phi}_1^i(e)|<<1$ when $t$ becomes very small.
Hence for small $t$ one has
 \begin{equation}\label{Kgg}
\langle\tilde{\underline{\Psi}}^t_{\underline{g}_2(e)},\tilde{\underline{\Psi}}^t_{\underline{g}_1(e)}\rangle \simeq
e^{\underline{K}(\underline{g}_2(e), \underline{g}_1(e))/t},
\end{equation}
 where
\begin{eqnarray}
&&\underline{K}(\underline{g}_2(e), \underline{g}_1(e))\\\nonumber
&=& \sum_{i\in\{1,2,3\}}[-\frac{\frac{1}{2}(\underline{\phi}_2^i(e)-\underline{\phi}_1^i(e))^2}{2} -\frac{\frac{1}{2}(\underline{X}_2^i(e)-\underline{X}_1^i(e))^2}{2} +\frac{\mathbf{i}(\underline{\phi}_2^i(e)-\underline{\phi}_1^i(e))(\underline{X}_2^i(e)+\underline{X}_1^i(e))}{2}]\\\nonumber
&=&\sum_{i\in\{1,2,3\}}[(\frac{\mathbf{i}\underline{\phi}_2^i(e)+\underline{X}_2^i(e)}{2} -\frac{\mathbf{i}\underline{\phi}_1^i(e)-\underline{X}_1^i(e)}{2} )^2-\frac{(\underline{X}_2^i(e))^2}{2}-\frac{(\underline{X}_1^i(e))^2}{2}].
\end{eqnarray}

For simplicity, we consider topological simple graphs $\gamma$ such as the cubic graph and focus on the transition amplitude $\underline{A}_{\underline{g},\underline{g}'}$ defined by the non-graph-changing physical Hamiltonian $\hat{\underline{\mathbf{H}}}$ as
\begin{equation}
\underline{A}_{\underline{g},\underline{g}'}:=\langle\underline{\Psi}_{\underline{g}}^t|U(T)|\underline{\Psi}_{\underline{g}'}^t\rangle,\quad \text{with}\  U(T):=\exp(-\frac{\mathbf{i}}{\hbar}T\hat{\underline{\mathbf{H}}}),
\end{equation}
where $\underline{\Psi}_{\underline{g}}^t(\underline{h})=\prod_{e\in\gamma}\underline{\Psi}^t_{\underline{g}(e)}(\underline{h}(e))$, $\underline{g}=\{\underline{g}(e)=e^{\mathbf{i}\sum_{i}(\underline{\phi}^i(e)+\mathbf{i}\underline{X}^i(e))}\}_{e\in\gamma}$ and $\underline{h}=\{\underline{h}(e)=e^{\mathbf{i}\sum_{i}(\underline{\theta}^i(e))}\}_{e\in\gamma}$.
Following the standard coherent state functional integral method, we discretize the time $T$ into $N$ steps, where $N$ can be arbitrarily large, thus that each step $\Delta \tau=T/N$ is arbitrarily small. Then the amplitude $\underline{A}_{\underline{g},\underline{g}'}$ can be written as a discrete path integral with an effective action $\underline{S}[\underline{g}]$ by the approximation \eqref{Kgg}:
\begin{equation}
\underline{A}_{\underline{g},\underline{g}'}=||\underline{\Psi}_g^t||||\underline{\Psi}_{g'}^t||\int\prod_{\imath=1}^{N+1}dg_\imath e^{\underline{S}[\underline{g}]/t},
\end{equation}
where the effective action is given by
\begin{equation}\label{action2}
\underline{S}[\underline{g}]=\sum_{\imath=0}^{N+1}\underline{K}(\underline{g}_{\imath+1}, \underline{g}_\imath)-\frac{\mathbf{i}\kappa}{a^2}\sum_{\imath=1}^{N}\Delta \tau[\frac{\langle\underline{\Psi}^t_{\underline{g}_{\imath+1}}|\hat{\mathbf{H}}|{\underline{\Psi}}^t_{\underline{g}_{i}}\rangle }{\langle\underline{\Psi}^t_{\underline{g}_{\imath+1}}|{\underline{\Psi}}^t_{\underline{g}_{i}}\rangle }+\mathbf{i}\tilde{\varepsilon}_{\imath+1,\imath}(\frac{\Delta \tau}{\hbar})],\quad \underline{g}_0=\underline{g}', \underline{g}_{N+2}=\underline{g},
\end{equation}
with
\begin{equation}
\underline{K}(\underline{g}_{\imath+1}, \underline{g}_\imath) =\sum_{e\in\gamma}\sum_{i\in\{1,2,3\}}[-(\frac{\underline{\phi}_{\imath+1}^i(e)-\mathbf{i}\underline{X}_{\imath+1}^i(e)}{2} -\frac{\underline{\phi}_\imath^i(e)+\mathbf{i}\underline{X}^i_\imath(e)}{2} )^2+\frac{(\mathbf{i}\underline{X}_{\imath+1}^i(e))^2}{2}+\frac{(\mathbf{i}\underline{X}_\imath^i(e))^2}{2}],
\end{equation}
and $\tilde{\varepsilon}_{\imath+1,\imath}(\frac{\Delta \tau}{\hbar})$ satisfying $\lim_{\Delta\tau\rightarrow0}\tilde{\varepsilon}_{\imath+1,\imath}(\frac{\Delta \tau}{\hbar})=0$.

Denoting $\underline{g}_\imath^{\varepsilon}(e)=\underline{g}_\imath(e)e^{\mathbf{i}\sum_{i}\varepsilon_\imath^i}, \text{for}\ \imath=1,...,N$,  the variations of the action \eqref{action2} with respect to $\varepsilon_\imath^i$ and their complex conjugate $\bar{\varepsilon}_\imath^i$ can give the EOMs.
 For $\imath = 1,...,N$, the variation with respect to $\varepsilon_{\imath}^i(e)$ gives
\begin{equation}\label{EOM1} \frac{\underline{\phi}_{\imath+1}^i(e)-\mathbf{i}\underline{X}_{\imath+1}^i(e)}{2} -\frac{\underline{\phi}^i_{\imath}(e)-\mathbf{i}\underline{X}^i_{\imath}(e)}{2} = \frac{\mathbf{i}\kappa}{a^2}\Delta \tau\frac{\delta}{\delta\varepsilon_{\imath}^i(e)}\left[\frac{\langle\underline{\Psi}^t_{\underline{g}_{\imath+1}}| \hat{\underline{\mathbf{H}}}|{\underline{\Psi}}^t_{\underline{g}^\varepsilon_{\imath}}\rangle }{\langle\underline{\Psi}^t_{\underline{g}_{\imath+1}}|{\underline{\Psi}}^t_{\underline{g}^\varepsilon_{\imath}}\rangle }\right]_{\vec{\varepsilon}=0}.
\end{equation}
For $\imath=N+1$, the variation with respect to $\varepsilon^i_{N+1}(e)$ gives
\begin{equation}\label{EOM2} \frac{\underline{\phi}_{N+2}^i(e)-\mathbf{i}\underline{X}_{N+2}^i(e)}{2} -\frac{\underline{\phi}_{N+1}^i(e)-\mathbf{i}\underline{X}_{N+1}^i(e)}{2} = 0.
\end{equation}
 For $\imath = 2,...,N+1$, the variation with respect to $\bar{\varepsilon}_\imath^i(e)$ gives
\begin{equation}\label{EOM3} -\frac{\underline{\phi}^{i}_{\imath}(e)+\mathbf{i}\underline{X}^{i}_{\imath}(e)}{2} +\frac{\underline{\phi}_{\imath-1}^i(e)+\mathbf{i}\underline{X}_{\imath-1}^i(e)}{2} = \frac{\mathbf{i}\kappa}{a^2}\Delta \tau\frac{\delta}{\delta\bar{\varepsilon}_\imath^i(e)} \left[\frac{\langle\underline{\Psi}^t_{\underline{g}^\varepsilon_{\imath}}|\hat{\underline{\mathbf{H}}} |{\underline{\Psi}}^t_{\underline{g}_{\imath-1}}\rangle }{\langle\underline{\Psi}^t_{\underline{g}^\varepsilon_{\imath}}|{\underline{\Psi}}^t_{\underline{g}_{\imath-1}}\rangle }\right]_{\vec{\varepsilon}=0}.
\end{equation}
For $\imath=1$, the variation with respect to $\bar{\varepsilon}^i_1(e)$ gives
\begin{equation}\label{EOM4}
 \frac{\underline{\phi}_{1}^i(e)+\mathbf{i}\underline{X}_{1}^i(e)}{2} -\frac{\underline{\phi}_{0}^i(e)-\mathbf{i}\underline{X}_{0}^i(e)}{2} = 0.
\end{equation}
We can approximate solutions of EOMs in the continuum limit (in the time direction) as $\Delta \tau\rightarrow 0$. This leads to $\underline{g}_\imath\rightarrow \underline{g}_{\imath+1}$. In this limit, the matrix elements of $\hat{\underline{\mathbf{H}}}$ in the right-hand sides of Eqs.\eqref{EOM1} and \eqref{EOM3} reduce to the expectation value of $\hat{\underline{\mathbf{H}}}$ as follows.\\
\textbf{Lemma 1}
\begin{equation}
\lim_{\underline{g}_i\rightarrow \underline{g}_{i+1}\equiv \underline{g}}\frac{\partial}{\partial \varepsilon^i_\imath(e)}\left[\frac{\langle \underline{\Psi}^t_{\underline{g}_{\imath+1}}|\hat{\underline{\mathbf{H}}}|\underline{\Psi}^t_{\underline{g}_{\imath}^\varepsilon}\rangle}{\langle \underline{\Psi}^t_{\underline{g}_{\imath+1}}|\underline{\Psi}^t_{\underline{g}_{\imath}^\varepsilon}\rangle}\right]_{\vec{\varepsilon}=0} =\left.\frac{\partial \langle \tilde{\underline{\Psi}}^t_{\underline{g}^\varepsilon}|\hat{\underline{\mathbf{H}}}|\tilde{\underline{\Psi}}^t_{\underline{g}^\varepsilon}\rangle}{\partial \varepsilon^i(e)}\right|_{\vec{\varepsilon}=0},
\end{equation}
\begin{equation}
\lim_{\underline{g}_{\imath-1}\rightarrow \underline{g}_{\imath}\equiv \underline{g}}\frac{\partial}{\partial \bar{\varepsilon}_\imath^i(e)}\left[\frac{\langle \underline{\Psi}^t_{\underline{g}^\varepsilon_{\imath}}|\hat{\underline{\mathbf{H}}}|\underline{\Psi}^t_{\underline{g}_{\imath-1}}\rangle}{\langle \underline{\Psi}^t_{\underline{g}^\varepsilon_{\imath}}|\underline{\Psi}^t_{\underline{g}_{\imath-1}}\rangle}\right]_{\vec{\varepsilon}=0} =\left.\frac{\partial \langle \tilde{\underline{\Psi}}^t_{\underline{g}^\varepsilon}|\hat{\underline{\mathbf{H}}}|\tilde{\underline{\Psi}}^t_{\underline{g}^\varepsilon}\rangle}{\partial \bar{\varepsilon}^i(e)}\right|_{\vec{\varepsilon}=0}.
\end{equation}
Similar to the case of Lemma 4.2 in Ref.\cite{Han_2020}, this lemma can be proved based on the following identities
\begin{equation}
\frac{\partial }{\partial \varepsilon^i(e)}\langle {\underline{\Psi}}^t_{\underline{g}^\varepsilon}|\hat{\underline{\mathbf{H}}}|{\underline{\Psi}}^t_{\underline{g}^\varepsilon}\rangle =\frac{\partial }{\partial \varepsilon^i(e)}\int d\underline{h}\overline{(\hat{\mathbf{H}}^{\dagger} {\underline{\Psi}}^t_{\underline{g}^\varepsilon})(\underline{h})}{\underline{\Psi}}^t_{\underline{g}^\varepsilon}(\underline{h}) =\int d\underline{h}\overline{(\hat{\underline{\mathbf{H}}}^{\dagger} {\underline{\Psi}}^t_{\underline{g}^\varepsilon})(\underline{h})}\frac{\partial }{\partial \varepsilon^i(e)}{\underline{\Psi}}^t_{\underline{g}^\varepsilon}(\underline{h})
\end{equation}
where the integral is taken over a compact space and  $\overline{(\hat{\mathbf{H}}^{\dagger} {\phi}^t_{\underline{g}^\varepsilon})(\underline{h})}$ depends on $\varepsilon^i(e)$ anti-holomorphically, and
\begin{equation}
\frac{\partial}{\partial {\varepsilon}^i(e)}\left. \underline{\Psi}^t_{\underline{g}^{\varepsilon}}(\underline{h})\right|_{\vec{\varepsilon}=0}=-\hat{V}_e^i\underline{\Psi}^t_{\underline{g}}(\underline{h})
\end{equation}
with $\hat{V}_e^i$ being the vector field on $U(1)^3$ defined by $\hat{V}^if(\underline{h})=\left.\frac{d}{d\varepsilon^i}\right|_{\vec{\varepsilon}=0}f(\underline{h}e^{\mathbf{i}\varepsilon^i})$.

 Lemma 1 implies that the EOMs with continuous time $\tau$ involve only the expectation value of $\hat{\underline{\mathbf{H}}}$. We assume that $\hat{\underline{\mathbf{H}}}$ has the correct semiclassical limit, in the sense that $\langle {\underline{\Psi}}^t_{\underline{g}^\varepsilon}|\hat{\underline{\mathbf{H}}}|{\underline{\Psi}}^t_{\underline{g}^\varepsilon}\rangle$ can reproduce the classical Hamiltonian $\underline{\mathbf{H}}$ as a function on the $U(1)^3$ holonomy-flux phase space in the semiclassical limit as:
\begin{equation}\label{semiclassicality2}
\langle {\underline{\Psi}}^t_{\underline{g}^\varepsilon}|\hat{\underline{\mathbf{H}}}| {\underline{\Psi}}^t_{\underline{g}^\varepsilon}\rangle=\underline{\mathbf{H}}[\underline{g}^\varepsilon]+\mathcal{O}(t).
\end{equation}
Notice that the (non-graph-changing) physical Hamiltonian $\hat{\underline{\mathbf{H}}}$ is just a combination of the basic $U(1)^3$ holonomy and flux operators. Thus it is reasonable to assume that Eq. \eqref{semiclassicality2} holds based on the simple Gaussian damping formulation and the Peakedness properties of the $U(1)^3$ coherent states \cite{2001Gauge,2000Gauge}. Therefore, all of the expectation value $\langle {\underline{\Psi}}^t_{\underline{g}^\varepsilon}|\hat{\underline{\mathbf{H}}}| {\underline{\Psi}}^t_{\underline{g}^\varepsilon}\rangle$ can be replaced by the classical Hamiltonian $\underline{\mathbf{H}}[\underline{g}^\varepsilon]$ in EOMs by taking $t\rightarrow0$. Then, the effective action in the time continuous limit, $\underline{\mathcal{S}}[\underline{g}]=\lim_{\Delta\tau\rightarrow0}\underline{S}[\underline{g}]$, reads
\begin{equation}
\underline{\mathcal{S}}[\underline{g}]=\mathbf{i}\int_{0}^Td\tau\left(\sum_{e\in\gamma}\sum_{i\in\{1,2,3\}} \underline{X}_e^i\frac{d\underline{\phi}^i_e}{d\tau}-\frac{\kappa}{a^2}\underline{\mathbf{H}}[\underline{g}]+\mathcal{O}(t) \right).
\end{equation}
The inherent Poisson algebra of this effective action is
\begin{equation}\label{effcPoisson2}
\{\underline{\phi}^i_e,\underline{\phi}^j_{e'}\}=\{\underline{X}^i_e,\underline{X}^j_{e'}\}=0,\quad \{\underline{\phi}^i_e,\underline{X}^j_{e'}\}=\frac{\kappa}{a^2}\delta_{e,e'}\delta^{ij}.
\end{equation}
It is easy to see that this algebra is equivalent to the original Poisson algebra \eqref{originalU(1)3} for $U(1)^3$ LQG.
 The corresponding EOMs can be reduced to the following form:
\begin{eqnarray}\label{EOMU13}
&&\frac{d\underline{X}_e^i}{d\tau}=-\frac{\kappa}{a^2}\frac{\partial\underline{\mathbf{H}}[\underline{g}]}{\partial \underline{\phi}_e^i},\\\nonumber
&&\frac{d \underline{\phi}_e^i}{d\tau}=\frac{\kappa}{a^2}\frac{\partial\underline{\mathbf{H}}[\underline{g}]}{\partial \underline{X}_e^i},
\end{eqnarray}
in the limits $\Delta\tau\rightarrow0$ and $t\rightarrow0$.
\subsection{Comparison with the $SU(2)$ LQG}
To compare the weak coupling limit of the effective dynamics of the $SU(2)$ LQG with that of $U(1)^3$ LQG, we first recall the relation between the basic variables in these two theories. Firstly, the reparametrization \eqref{repa} implies that
in the weak coupling limit of small $\underline{\phi}^i_e$, one has $\phi_e^i=\underline{\phi}_e^i$ and $X_e^i=\underline{X}_e^i$ up to higher order terms. Thus the $\underline{\phi}_e^i$ of $U(1)^3$ holonomy can parametrize the ${\phi}_e^i$ of $SU(2)$ holonomy in the weak coupling limit. Secondly, Eq.\eqref{hp33} implies that the Poisson brackets among $(\phi_e^i,p_e^i,\tilde{p}_e^i)$ are consistent with those of $(\underline{\phi}_e^i,\underline{p}_e^i,\underline{\tilde{p}}_e^i)$ in the weak coupling limit.
Since ${\mathbf{H}}[{g}]$ (or $\underline{\mathbf{H}}[\underline{g}]$) are functions of $(\phi_e^i,p_e^i,\tilde{p}_e^i)$ (or $(\underline{\phi}_e^i,\underline{p}_e^i,\underline{\tilde{p}}_e^i)$) and their Poisson brackets, we can immediately have the relation
\begin{eqnarray}\label{relation}
&&{\mathbf{H}}[{g}]=\underline{\mathbf{H}}[\underline{g}]
\end{eqnarray}
at the weak coupling limit based on the reparametrization \eqref{repa}. Further, we note that
\begin{equation}\label{pXpX}
\frac{\delta  p_e^i}{\delta  X_e^i}=\frac{\delta \underline{ p}_e^i}{\delta  \underline{X}_e^i}=-1,\quad \frac{\delta  \tilde{p}_e^i}{\delta {X}_e^i}=\frac{\delta \underline{ \tilde{p}}_e^i}{\delta  \underline{{X}}_e^i}=1,
\end{equation}
at the weak coupling limit. Hence, Eqs. $\eqref{relation}$, \eqref{pXpX} and the reparametrization \eqref{repa} ensure that the effective EOMs \eqref{EOMU13} in $U(1)^3$ LQG is consistent with the effective EOMs \eqref{EOMo} in $SU(2)$ LQG in the weak coupling limit up to higher order corrections of $t$. This consistent result can be used to deal with the ``Gauss'' constraint (closure condition) in the weak coupling $U(1)^3$ LQG, which is neglected in the above discussion. Notice that the Gaussian constraint in the effective dynamics of $SU(2)$ LQG can be satisfied by the corresponding constraint on the labelling parameters of the boundary coherent state \cite{Han_2020}, and the effective dynamics preserves the constraint. Thus, the consistency between the effective EOMs of  $U(1)^3$ and $SU(2)$ LQG in weak coupling limit implies that the ``Gauss'' constraint can also be implemented in the effective dynamics of the weak coupling $U(1)^3$ LQG.
 \section{Conclusion and Discussion}
   To summarize, in order to relate the holonomy-flux algebra of $SU(2)$ LQG to that of $U(1)^3$ LQG, a parametrization of $SU(2)$ holonomy-flux variables by $U(1)^3$ holonomy-flux variables is constructed as Eqs.\eqref{repa}. It is shown that the $SU(2)$ holonomy-flux algebra can be reproduced in the $U(1)^3$ holonomy-flux phase space based on this parametrization in the weak coupling limit. Thus, the $U(1)^3$ holonomy-flux variables can be endowed with certain specific geometric meaning of $SU(2)$ holonomy-flux variables  in the weak coupling limit. With this re-parametrization, the Hamiltonian constraint in the weak coupling $U(1)^3$ LQG is introduced by replacing the $SU(2)$ holonomy-flux variables in the Hamiltonian constraint of the $SU(2)$ LQG with the corresponding reparametrization variables in the $U(1)^3$ holonomy-flux phase space.
Based on this Hamiltonian, the effective dynamics is derived from the coherent state path integral of the weak coupling $U(1)^3$ LQG. It is shown that the effective EOMs obtained are consistent with those of $SU(2)$ LQG in the weak coupling limit, provided that the expectation values of the Hamiltonian operators with respect to the coherent states in these two theories coincide with the corresponding classical Hamiltonians respectively. Then, in the weak coupling limit, the $U(1)^3$ LQG could reflect the main characters of the $SU(2)$ LQG at both kinematics and effective dynamics levels.

Several interesting issues deserve  further investigating based on the effective dynamics of weak coupling $U(1)^3$ LQG. First, as a manageable dynamical model of quantum spacetime, the weak coupling $U(1)^3$ LQG can be coupled with matter fields. Especially, the coupled model can be used to study whether QFT on curved spacetimes could be obtained as certain semiclassical limit of LQG \cite{Sahlmann:2002qj}. Second, it is expected to extend this weak coupling method for $SU(2)$ LQG in (1+3)-dimensions to higher dimensional LQG \cite{Bodendorfer:Ha,Bodendorfer:Qu,PhysRevD.103.086016}. Note that the $SO(D+1)$ LQG in higher dimensions has much complicated gauge groups and an anomalous simplicity constraint \cite{long2019coherent,long2020operators,Long:2020euh}. Hence the explicit calculations for the dynamics become extremely hard, even at the level of effective dynamics. Nevertheless, it is very possible that the much simpler weak coupling LQG with Abelian gauge group (e.g., $U(1)^D$) in higher dimensions may also reflect some key characters of the full $SO(D+1)$ LQG in proper limit.
\section*{Acknowledgments}
This work is supported by the National Natural Science Foundation of China (NSFC) with Grants No. 12047519, No. 11775082, No. 11875006 and No. 11961131013. G.L. acknowledges the support by China Postdoctoral Science Foundation with Grant No. 2021M691072.

\bibliographystyle{unsrt}

\bibliography{ref}


\end{document}